\documentclass[fleqn,usenatbib]{mnras}

\usepackage{newtxtext,newtxmath}

\usepackage[T1]{fontenc}
\usepackage{ae,aecompl}

\usepackage{graphicx}	
\usepackage{subcaption} 
\usepackage{booktabs}
\usepackage{multirow}
\usepackage{cleveref}
\usepackage{rotating}

\title[Informing 21-cm antenna design via simulation]{Informing antenna design for sky-averaged 21-cm experiments using a simulated Bayesian data analysis pipeline}

\author[D. Anstey et al.]{
Dominic Anstey,$^{1}$\thanks{E-mail: \href{da401@mrao.cam.ac.uk}{da401@mrao.cam.ac.uk}}
John Cumner,$^{1}$\thanks{E-mail: \href{jmc227@mrao.cam.ac.uk}{jmc227@mrao.cam.ac.uk}}
Eloy de Lera Acedo$^{1}$\thanks{E-mail: \href{eloy@mrao.cam.ac.uk}{eloy@mrao.cam.ac.uk}}
 and Will Handley$^{1,2}$\thanks{E-mail: \href{wh260@mrao.cam.ac.uk
}{wh260@mrao.cam.ac.uk}}
\\
$^{1}$Astrophysics Group, Cavendish Laboratory, J. J. Thomson Avenue, Cambridge, CB3 0HE, UK\\
$^{2}$Kavli Institute for Cosmology, Madingley Road, Cambridge, CB3 0HA, UK\\
}

\date{Accepted XXX. Received YYY; in original form ZZZ}

\pubyear{2020}

\begin{document}
\label{firstpage}
\pagerange{\pageref{firstpage}--\pageref{lastpage}}
\maketitle

\begin{abstract}
Global 21-cm experiments aim to measure the sky averaged HI absorption signal from cosmic dawn and the epoch of reionisation. However, antenna chromaticity coupling to bright foregrounds can introduce distortions into the observational data of such experiments. We demonstrate a method for guiding the antenna design of a global experiment through data analysis simulations. This is done by performing simulated observations for a range of inserted 21-cm signals, then attempting to identify the signals with a data analysis pipeline. We demonstrate this method on five antennae that were considered as potential designs for the Radio Experiment for the Analysis of Cosmic Hydrogen (REACH); a conical log spiral antenna, an inverted conical sinuous antenna and polygonal-, rectangular- and elliptical-bladed dipoles. We find that the log spiral performs significantly better than the other antennae tested, able to correctly and confidently identify every inserted 21-cm signal. In second place is the polygonal dipole antenna, which was only unable to detect signals with both very low amplitudes of $0.05$\,K and low centre frequency of $80$\,MHz. The conical sinuous antenna was found to perform least accurately, only able to detect the highest amplitude 21-cm signals, and even then with biases. We also demonstrate that, due to the non-trivial nature of chromatic distortion and the processes of correcting for it, these are not the results that could have been expected superficially from the extent of chromatic variation in each antenna.
\end{abstract}




\section{Introduction}\label{sec:intro}
There are few reliable mechanisms available to probe the cosmic history of the dark ages between the Epoch of Recombination, when the Cosmic Microwave Background (CMB) was emitted, and the formation of modern astrophysical structures. One of the most promising is the use of the HI 21-cm absorption line. The principle behind this method is that neutral hydrogen, which is the predominant component of the intergalactic medium, has a hyperfine absorption line at 21\,cm. Therefore, it will absorb (emit) at this wavelength when its spin temperature is less (greater) than the background radiation temperature. During cosmic dawn and the epoch of reionisation, many effects are predicted to occur that will alter the spin temperature of the surrounding neutral hydrogen. These include the Wouthuysen-Field effect \citep{wouthuysen52,field58} from stellar Lyman-$\alpha$ radiation, X-ray heating and the reduction of the neutral hydrogen fraction due to reionisation \citep{furlanetto16}. These effects together are predicted to result in a small but distinctive dip, and possibly a subsequent peak, in the background radiation temperature at around $\sim100\mathrm{\,MHz}$.

The aim of 21-cm cosmology is to measure and study this signal. There are two main approaches to this being taken. The first is through interferometric experiments such as HERA \citep{deboer17}, the SKA \citep{dewdney09}, PAPER \citep{parsons10}, LOFAR \citep{vanhaarlem13} and the MWA \citep{lonsdale09}. These experiments aim to map the full power spectra of hydrogen 21-cm absorption. The second is to use a small number of broad-beam, wideband antennae to measure the sky-averaged, `global' 21-cm signal. These have the advantages of being much simpler to build than large, interferometric experiments and that they require much shorter integration times to achieve low enough noise to detect the signal. However, they are also very difficult to calibrate. Such `global' 21-cm experiments include EDGES \citep{bowman18}, SARAS \citep{singh18}, MIST (\url{http://www.physics.mcgill.ca/mist/}), PRIZM \citep{philip19}, SCI-HI \citep{voytek14}, LEDA \citep{price18}, DAPPER (\url{https://www.colorado.edu/ness/dark-ages-polarimeter-pathfinder-dapper}), CTP \citep{nhan18}, and REACH \citep{acedo19}. In addition, a combination of these two approaches could also be used, by attempting to measure the global signal with a closely packed interferometric array, such as in ASSASSIN \citep{mckinley20}.

The design of antennae used in these global experiments can vary significantly. For example, EDGES uses a rectangular-bladed dipole, whereas SARAS uses a spherical monopole antenna. A common design focus of many global experiments has been towards towards achieving as achromatic beams as possible.

In 2018, EDGES reported the first detection of the global 21-cm signal. The signal is much deeper than can be explained with current models, with a depth of $\sim0.5\mathrm{\,K}$. However, some studies, such as \citet{hills18}, \citet{bevins21}, \citet{sims20} and \citet{singh19}, have indicated that there may be uncorrected for systematic residuals in the data that may explain the anomalous depth. There is therefore a great deal of effort from other global experiments to perform an independent check of this result, with a particular focus on systematic residuals.

\subsection{Galactic Foregrounds}\label{sec:foregrounds}
One of the main difficulties in detecting the 21-cm cosmological signal is the presence of extremely bright radio foregrounds. These are predominately synchrotron and free-free radiation \citep{shaver99} and can reach many thousands of kelvin in brightness. This means they overwhelm the 21-cm signal, which has an expected amplitude of the order $\sim0.1-0.2$\,K, by several orders of magnitude. Being able to accurately separate these foregrounds from the signal is therefore vital to performing 21-cm cosmology.

One way of potentially performing this separation for global experiments is to exploit the spectral differences between the smooth, power law foregrounds and the non-smooth signal. This would involve fitting a smooth function, such as a log-polynomial, as in \cite{bowman18}, or a Maximally Smooth Function \citep{sathyanarayana15, bevins21}, to the observational data. Ideally, this would fit away the smooth foregrounds without fitting away much of the signal. However, this is made more difficult by the presence of non-smooth chromatic distortions.

\subsection{Antenna Chromaticity Effects}\label{sec:chrom_effects}
If observations are performed using an antenna with a pattern that varies with frequency, it will result in spatial structure coupling into the frequency domain. This introduces chromatic distortions into the data that are difficult to correct for. Compensating for this is extremely important for global 21-cm experiments, and so it has been studied extensively in the literature, for example in \citet{vedantham14}, \citet{bernadi15}, \citet{mozdzen16}, \citet{spinelli21}, \citet{mahesh21}, \citet{raghunathan21} and \citet{anstey21}.

Ideally, a global 21-cm experiment would employ an antenna that was perfectly achromatic in order to avoid this effect. However, this is difficult in practice. Since the 21-cm signal is expected to occur over a wide frequency range, a wideband antenna is required to measure it. A wider bandwidth will typically permit a greater number of current modes in the antenna, giving a greater potential for variation in the currents in the antenna induced by incident waves \citep{craeye14}. Therefore, chromatic variation in any given antenna will be higher for a wider band. Careful antenna design is therefore a vital component in designing a 21-cm experiment, to ensure the distortions introduced will not prevent the detection of the signal. The process of antenna design in the Radio Experiment for the Analysis of Cosmic Hydrogen (REACH) \citep{acedo19} consists of many steps. Further details beyond the technique discussed here can be found in Cumner et al. (in prep.).

In the REACH, a data analysis process has been developed to model the foregrounds of global 21-cm experiments, which is described in \cite{anstey21}. This process works by subdividing the sky into $N$ regions and assuming a constant spectral index in each to generate an approximate sky map, which can then be convolved with the beam of the antenna being used. By treating the spectral index of each region as a free parameter, this produces a parameterised model of the foregrounds that incorporates the chromatic distortions due to the antenna. The model can therefore fit away both foregrounds and chromatic distortions and allow a 21-cm signal to be detected in the data of a chromatic antenna.

However, this process cannot necessarily correct for the distortions of any arbitrarily chromatic antenna. It should also be noted that, for this pipeline, regular chromatic structures that are spectrally distinct from the 21-cm signal may aid in correctly identifying foregrounds, so attempting to design the antenna to be as achromatic as possible may not be the optimal case. Therefore, careful antenna design is still warranted. This paper proposes using end-to-end simulations through a data analysis pipeline as part of the design process of an experiment, rather than as a subsequent analysis.

In this paper, we consider five different designs for a global 21-cm experiment antenna and attempt to quantify how well each would perform. The paper will be ordered as follows. In \Cref{sec:antennae} we describe the five antennae we consider in this work and discuss what can and cannot be inferred from traditional methods of characterising chromaticity. In \Cref{sec:method}, we lay out a method of characterising a chromatic antenna's performance in a global 21-cm signal via physically motivated simulations. In \Cref{sec:results} we discuss the results of this analysis and its implications on informing antenna design in global 21-cm experiments and in \Cref{sec:conclusions}, we summarise our conclusions.

\section{Characterising Antenna Chromaticity}\label{sec:antennae}
\begin{figure*}
    \centering
    \begin{minipage}{0.25\textwidth}%
    \begin{subfigure}{\linewidth}%
        \includegraphics[width=\textwidth]{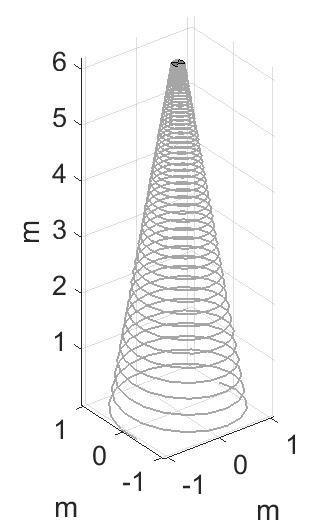}
         \caption{Conical Log Spiral Antenna}
         \label{subfig:log_spiral_image}
    \end{subfigure}
    \end{minipage}%
    \hspace{0.5mm}
    \begin{minipage}{0.7\textwidth}%
        \begin{minipage}{0.5\textwidth}%
            \begin{subfigure}{\linewidth}%
                 \includegraphics[width=\textwidth]{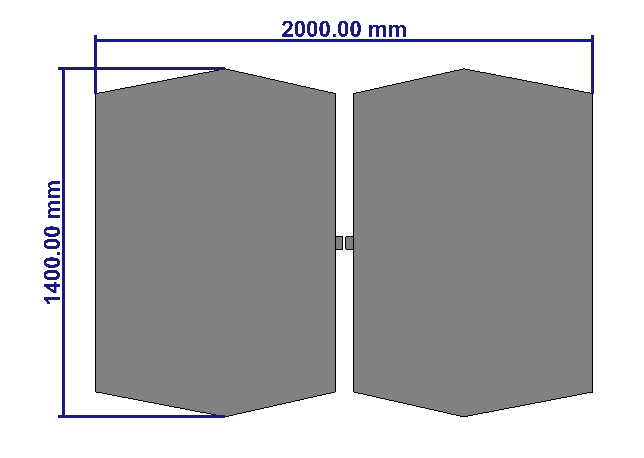}
                 \caption{Polygonal Dipole Antenna}
                 \label{subfig:polygonal_dipole_image}
            \end{subfigure}
        \end{minipage}
        \begin{minipage}{0.5\textwidth}%
            \begin{subfigure}{\linewidth}%
                 \includegraphics[width=\textwidth]{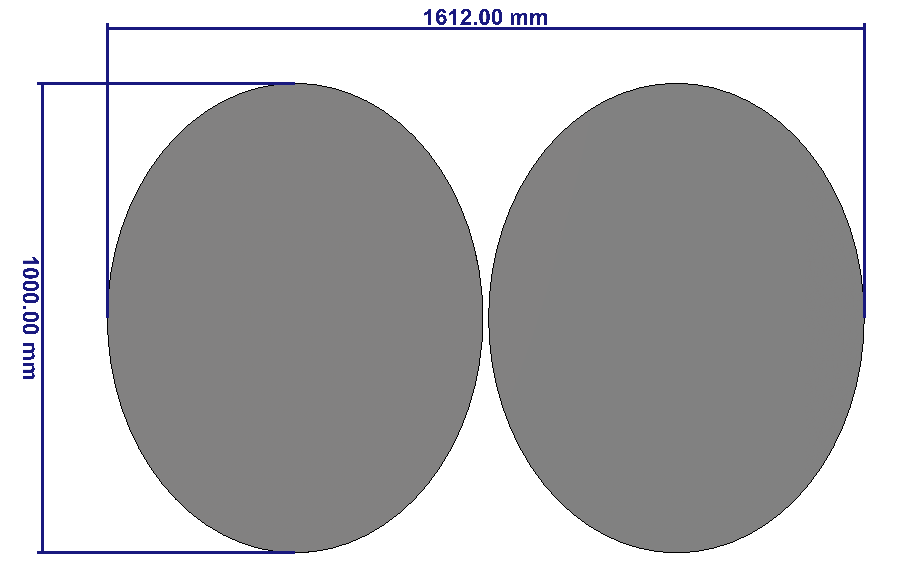}
                 \caption{Elliptical Dipole Antenna}
                 \label{subfig:elliptical_dipole_image}
            \end{subfigure}
        \end{minipage}
        \begin{minipage}{0.4\textwidth}%
            \begin{subfigure}{\linewidth}%
                 \includegraphics[width=\textwidth]{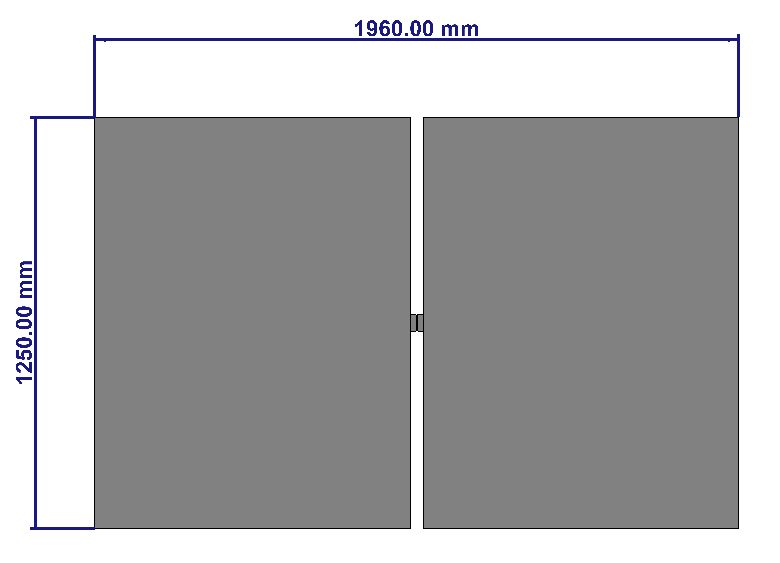}
                 \caption{Rectangular Dipole Antenna}
                 \label{subfig:square_dipole_image}
            \end{subfigure}
        \end{minipage}
        \hspace{12mm}
        \begin{minipage}{0.5\textwidth}%
            \begin{subfigure}{\linewidth}%
                 \includegraphics[width=\textwidth]{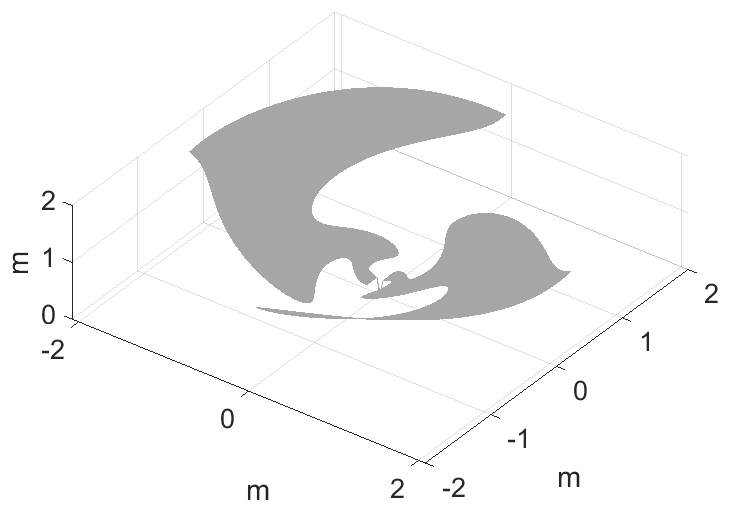}
                 \caption{Inverted Conical Sinuous Antenna}
                 \label{subfig:conical_sinuous_image}
            \end{subfigure}
        \end{minipage}
    \end{minipage}
        \caption{Diagrams of five antenna designs considered for use in REACH, which are analysed and compared in this paper to determine how effective each may be at detecting the HI 21-cm signal.}
        \label{fig:antenna_diagrams}
\end{figure*}

Dipoles, such as that used in \cite{bowman18}, are often considered for global 21-cm experiments due to their simplicity, which makes them easy to both model and construct. Therefore, we consider three different dipole antenna designs here. These are a rectangular-bladed dipole, an elliptical-bladed dipole and a polygonal-bladed dipole. We also consider two other very different antenna designs. These are a conical log spiral antenna \citep{dyson65} and an inverted conical sinuous antenna \citep{buck08}. These five antennae are all shown in \Cref{fig:antenna_diagrams}. 

For all the following work, we consider each antenna to be located in the Karoo Radio Reserve in South Africa, at -30.71131 degrees latitude and 21.4476236 degrees longitude. Each antenna has a 10m radius circular ground plane, not shown in \Cref{fig:antenna_diagrams}. For simplicity, soil effects are not considered in these simulations.

One way in which the chromaticity of an antenna can be characterised is 

\begin{equation}\label{eq:chrom_factor}
    A_\mathrm{chromaticity}\left(t,\nu\right) = \frac{\int^{4\pi}_{0}\left[T_\mathrm{sky}\left(t, \Omega,\nu_\mathrm{ref}\right)-T_\mathrm{CMB}\right]D\left(\Omega, \nu \right)\mathrm{d}\Omega}{\int^{4\pi}_{0}\left[T_\mathrm{sky}\left(t, \Omega,\nu_\mathrm{ref}\right)-T_\mathrm{CMB}\right]D\left(\Omega, \nu_\mathrm{ref}\right)\mathrm{d}\Omega},
\end{equation}
where $D\left(\Omega, \nu \right)$ is the antenna beam directivity, $\nu_\mathrm{ref}$ is a fixed reference frequency, $T_\mathrm{sky}\left(t, \Omega, \nu_\mathrm{ref}\right)$ is a sky map at that fixed reference frequency and $T_\mathrm{CMB}$ is the Cosmic Microwave Background temperature, taken to be 2.725\,K. This method is used in \cite{mozdzen17} and \cite{mozdzen18}. $T_\mathrm{CMB}$ is subtracted as its inclusion introduces slight errors to $A_\mathrm{chromaticity}\left(t,\nu\right)$. This effect is discussed in \cite{shen21}.

\begin{figure*}
     \centering
     \begin{subfigure}[b]{0.3\textwidth}
         \centering
         \includegraphics[width=\textwidth]{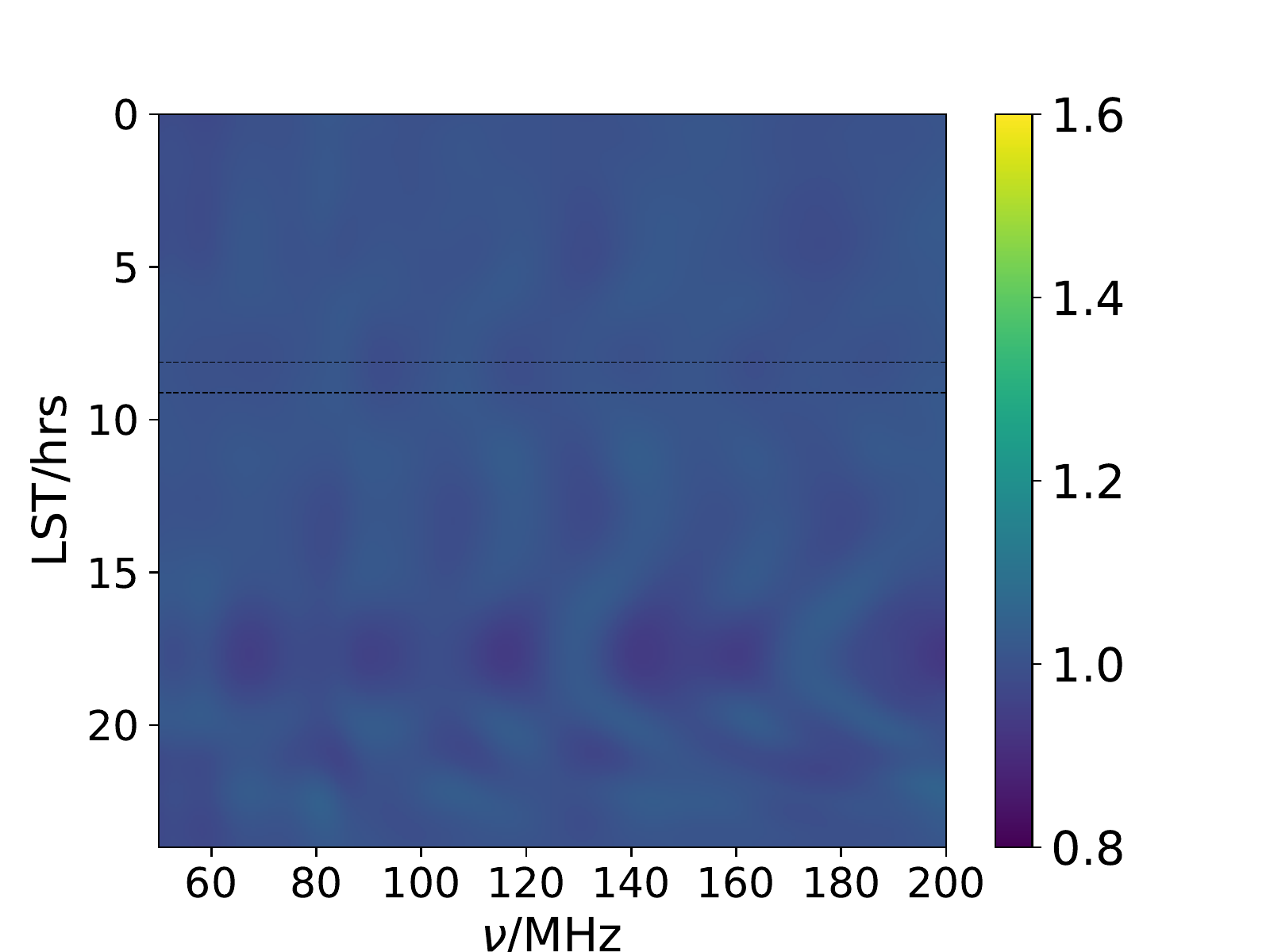}
         \caption{Log Spiral Antenna}
         \label{subfig:log_spiral_waterfall}
     \end{subfigure}
     \hfill
     \begin{subfigure}[b]{0.3\textwidth}
         \centering
         \includegraphics[width=\textwidth]{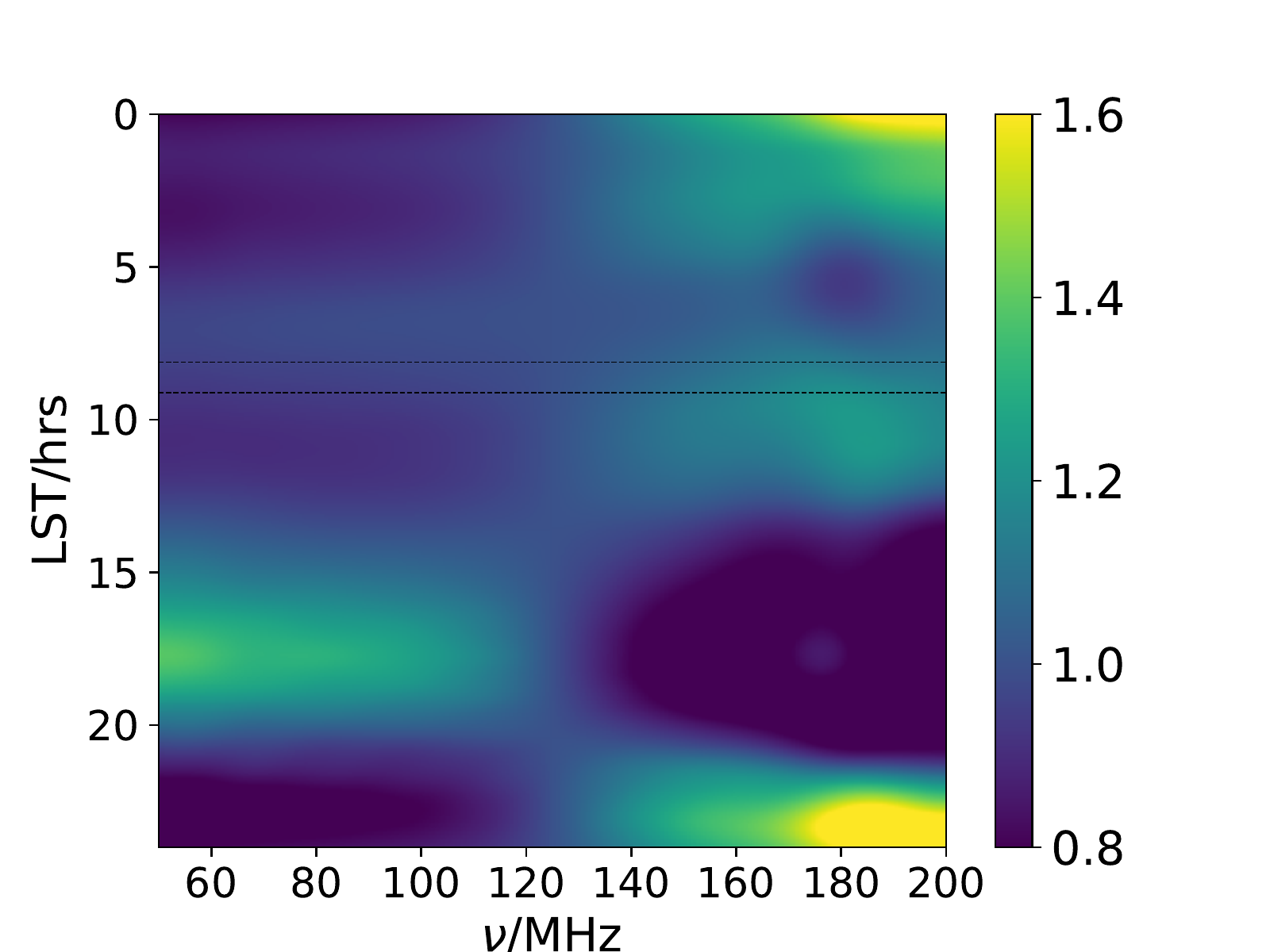}
         \caption{Polygonal Dipole Antenna}
         \label{subfig:polygonal_dipole_waterfall}
     \end{subfigure}
     \hfill
     \begin{subfigure}[b]{0.3\textwidth}
         \centering
         \includegraphics[width=\textwidth]{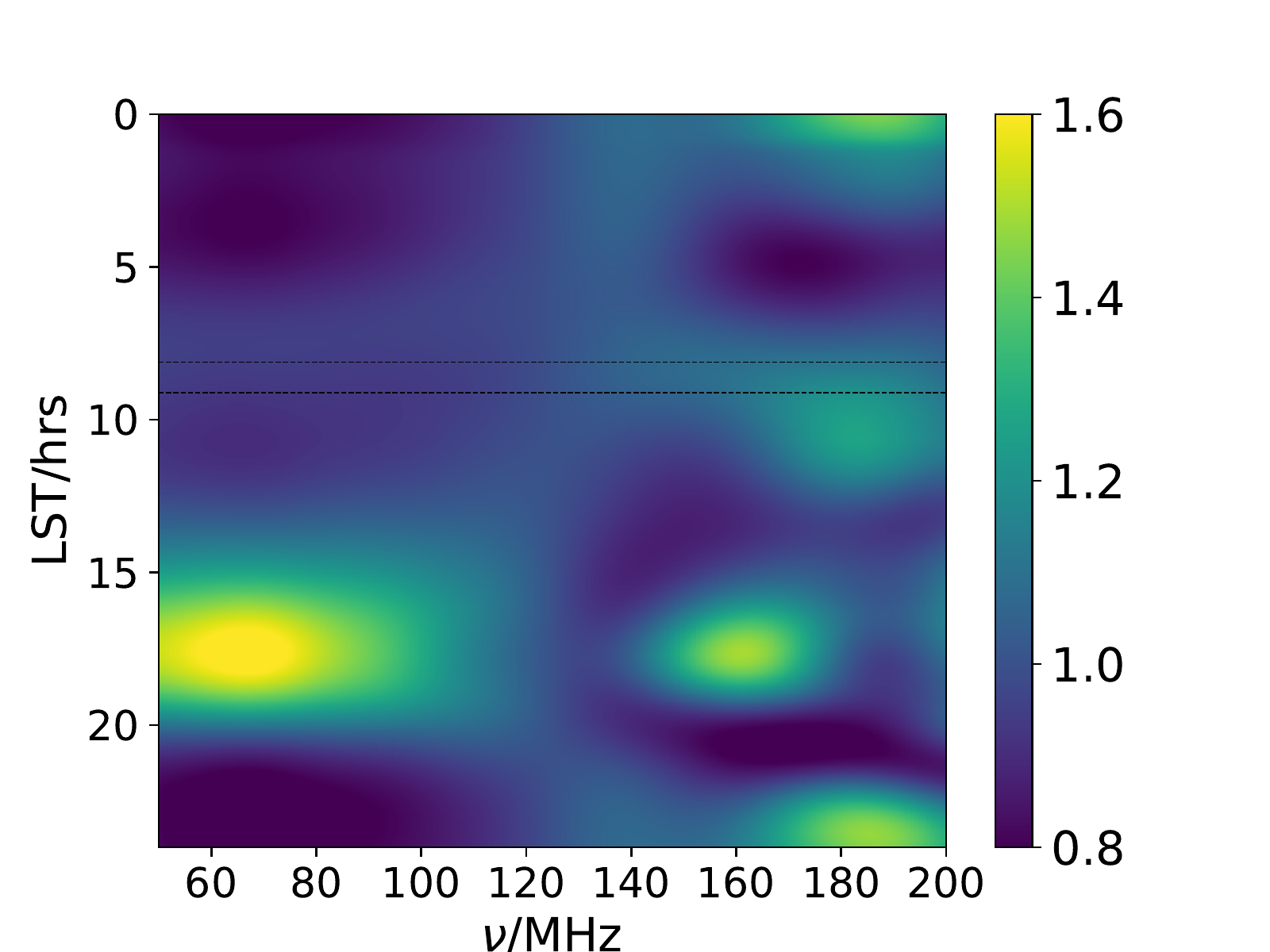}
         \caption{Elliptical Dipole Antenna}
         \label{subfig:elliptical_dipole_waterfall}
     \end{subfigure}
     \begin{subfigure}[b]{0.3\textwidth}
         \centering
         \includegraphics[width=\textwidth]{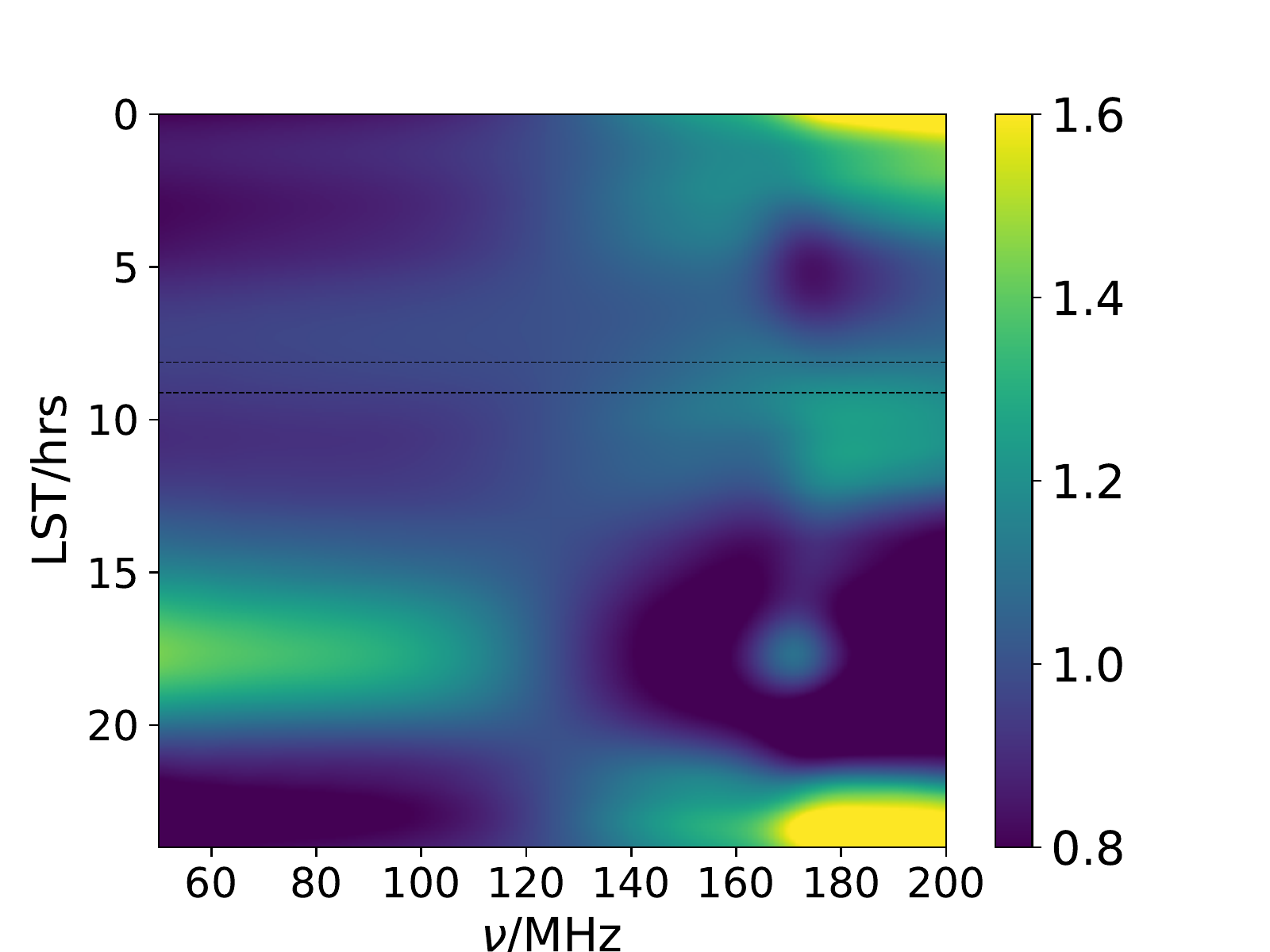}
         \caption{Rectangular Dipole Antenna}
         \label{subfig:square_dipole_waterfall}
     \end{subfigure}
     \begin{subfigure}[b]{0.3\textwidth}
         \centering
         \includegraphics[width=\textwidth]{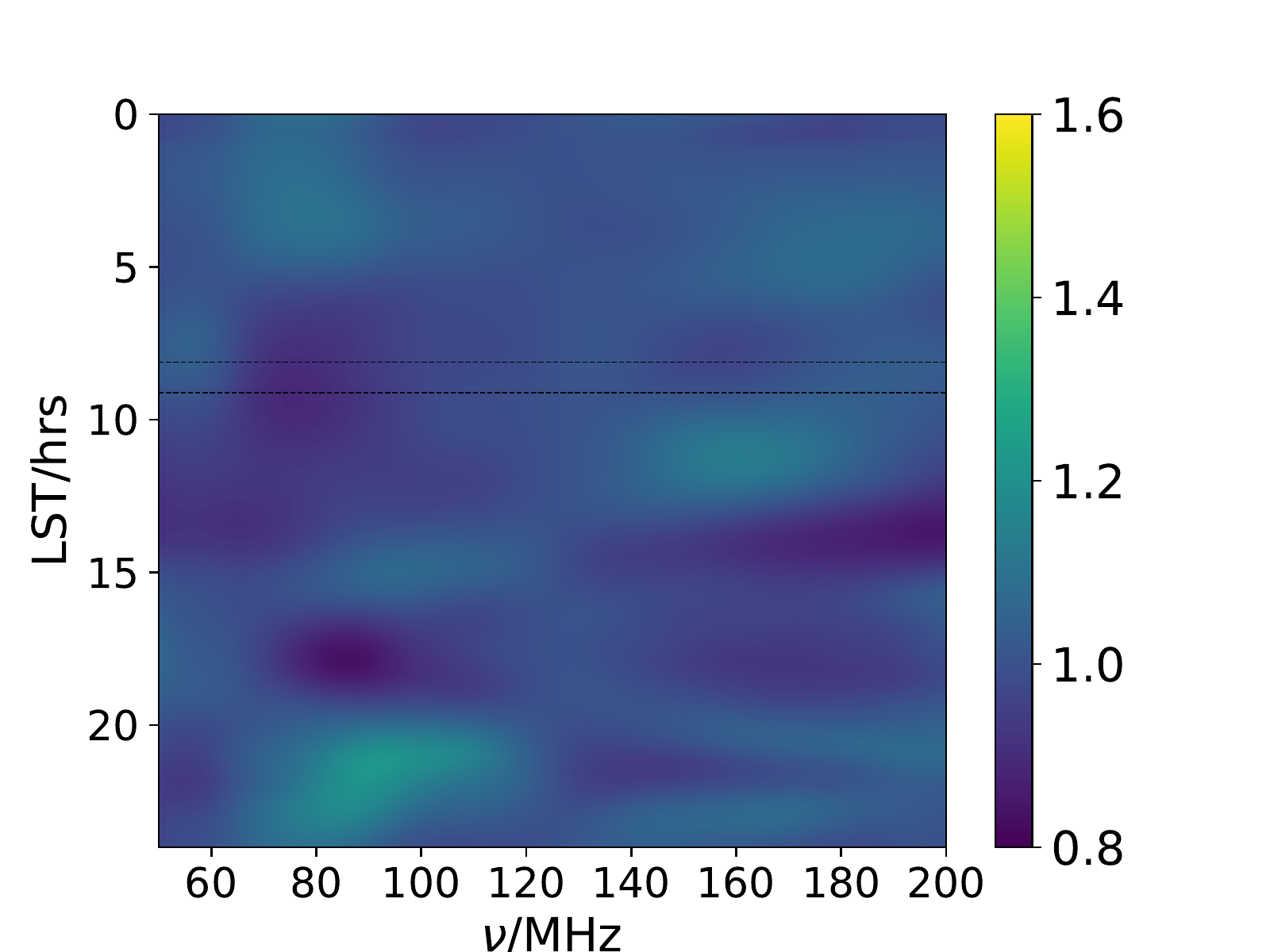}
         \caption{Conical Sinuous Antenna}
         \label{subfig:conical_sinuous_waterfall}
     \end{subfigure}
        \caption{Plots of the chromaticity factor $A_\mathrm{chromaticity}\left(t,\nu\right)$, given in \Cref{eq:chrom_factor}, for five antennae if they were to observe from the Karoo radio reserve. The LST range integrated over in this work is indicated by black lines.}
        \label{fig:waterfall_plots}
\end{figure*}

\Cref{fig:waterfall_plots} shows $A_\mathrm{chromaticity}\left(t,\nu\right)$ for each of the five antennae considered here over a full 24 hours, when observing from the Karoo radio reserve. The frequency range is taken as 50-200\,MHz, as this covers the full frequency band in which the global 21-cm signal is predicted to occur. In practice, these antennae may be unable to observe across the entirety of this band due to poor matching to the receiver at higher frequencies. However, for the purpose of this work, we are considering the effect of beam chromaticity only, so we will use the full 50-200\,MHZ band. We take $\nu_\mathrm{ref} = 125\mathrm{\,MHz}$, as this is the centre of the observing band. We also use an approximate sky model, as in \cite{mozdzen17} and \cite{mozdzen18}, of
\begin{equation}\label{eq:sky_T_scaling}
    T_\mathrm{sky}\left(t, \Omega, \nu_\mathrm{ref}\right) = \left[T_\mathrm{408}\left(t, \Omega\right) - T_\mathrm{CMB}\right]\left(\frac{\nu_\mathrm{ref}}{408}\right)^{-2.5} + T_\mathrm{CMB},
\end{equation}
where $T_\mathrm{408}\left(t, \Omega\right)$ is the reprocessed, destriped Haslam all-sky map \citep{remazeilles15}. 

The value of $A_\mathrm{chromaticity}\left(t,\nu\right)$, as defined in \Cref{eq:chrom_factor}, quantifies the variation in antenna temperature produced by that antenna, which is given by

\begin{equation}\label{eq:antenna_temp}
    T_\mathrm{A}\left(\nu, t\right) = \frac{1}{4\pi}\int_{0}^{4\pi}T_\mathrm{sky}\left(t, \Omega,\nu\right)D\left(\Omega, \nu \right)d\Omega,
\end{equation}
relative to the reference point. The version of $A_\mathrm{chromaticity}\left(t,\nu\right)$ used here is subject to the assumption that the spatial distribution of brightness temperature over the sky is the same across the entire band as it is at the reference frequency, changing only by an absolute scale factor. More detailed variations on this equation, such as that used in \citet{monsalve17}, are not subject to this.

If the extent of chromaticity is quantified by the range of the variation of $A_\mathrm{chromaticity}\left(t,\nu\right)$, it can be seen from \Cref{fig:waterfall_plots} that the log spiral antenna shows the least chromatic variation across the band. The greatest variation is shown by the rectangular bladed dipole.

However, none of these antennae are perfectly achromatic. All show chromatic variation to a greater or lesser degree. Therefore, in considering these designs for use in a global 21-cm experiment, it is important to identify what effect the chromaticity of each will have on attempts to detect the 21-cm signal, and how well it can be modelled or corrected for. This information cannot easily be determined just from a metric of the extent of the chromaticity such as $A_\mathrm{chromaticity}\left(t,\nu\right)$, as, although a perfectly achromatic antenna is ideal, antenna designs that attempt to minimise the chromaticity may not be optimal if it not made sufficiently small as to be negligible. For example, it may be easier to identify the 21-cm signal using a more chromatic antenna with a chromatic structure that is highly distinct from the signal, than with a less chromatic antenna that has chromatic structure more degenerate with it. Therefore, further analysis is needed to properly determine how effective a given antenna will be in a global 21-cm experiment.

\section{Method of characterising performance via end-to-end data simulation analysis }\label{sec:method}
We demonstrate a method of using end-to-end simulations as part of the design phase, to characterise potential antenna designs more thoroughly. Such simulations can be used to identify what 21-cm signals could be detected in data taken with a given antenna, and how accurately those signals are reconstructed. We perform this analysis as follows.

First, the antenna design being considered is modelled in an electromagnetic simulator and solver. CST microwave studio and FEKO were used here. From this, a normalised, frequency dependent directivity pattern of the antenna, $D\left(\Omega, \nu\right)$, can be produced.

Then, a simulation of an observation of the radio sky with this antenna can be performed, according to:

\begin{multline}\label{eq:data_sim}
    T_\mathrm{simulated}\left(\nu\right) = 
    \\\frac{1}{4\pi}\int^{4\pi}_{0}D\left(\Omega, \nu\right) \int^{t_\mathrm{end}}_{t_\mathrm{start}}T_\mathrm{simulated}\left(t, \Omega, \nu\right)dtd\Omega + \sigma_\mathrm{N},
\end{multline}
where $t_\mathrm{start}$ and $t_\mathrm{end}$ give the start and end times of the observation period of the simulation. For this work, we integrated for one hour, from $t_\mathrm{start}=8.12$ hours to $t_\mathrm{end} = 9.12$ hours, in LST. This was chosen as a moderate single-night integration period in which the galactic plane is not overhead and the variation of $A_\mathrm{chromaticity}\left(t,\nu\right)$ with frequency is low for all antennae. The integration time period is marked by black lines in \Cref{fig:waterfall_plots}. For this basic model, we used uncorrelated Gaussian noise, $\sigma_\mathrm{N}$, in the data, with a standard deviation of 0.025\,K.

\begin{figure}
 \includegraphics[width=\columnwidth]{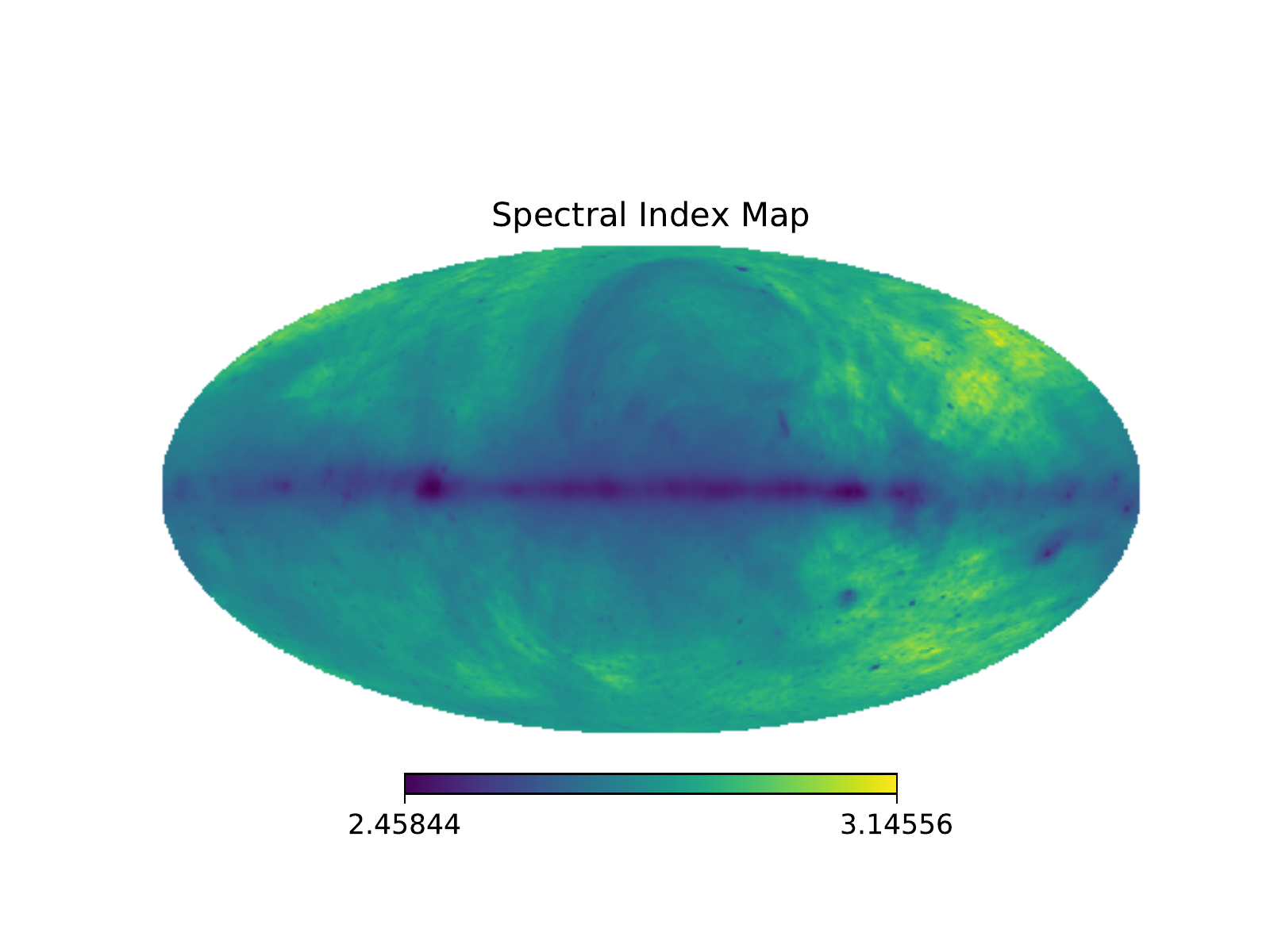}
 \caption{Spatially varying low-frequency spectral index map, in galactic coordinates, calculated by interpolating between two instances of the 2008 GMS \citep{deoliveiracosta08} at 230\,MHz and 408\,MHz according to \Cref{eq:B_map}.}
 \label{fig:B_map}
\end{figure}

$T_\mathrm{simulated}\left(t, \Omega, \nu\right)$ is a simulation of the entire sky in the observing band, for which we use: 
\begin{equation}
T_\mathrm{simulated}\left(t, \Omega, \nu\right) = \left(T_\mathrm{230}\left(\Omega, t\right) - T_\mathrm{CMB}\right)\left(\frac{\nu}{230}\right)^{-\beta\left(\Omega\right)} + T_\mathrm{CMB},
\end{equation}
where $\beta\left(\Omega\right)$ is a spatially varying spectral index map, shown in \Cref{fig:B_map}, which was calculated by interpolating between two instances of the 2008 Global Sky Model (GSM) \citep{deoliveiracosta08}, taken at 408\,MHz, $T_\mathrm{408}\left(\Omega\right)$, and 230\,MHz, $T_\mathrm{230}\left(\Omega\right)$, according to
\begin{equation}\label{eq:B_map}
    \beta\left(\Omega\right) = \frac{\log{\left(\frac{T_\mathrm{230}\left(\Omega\right)-T_\mathrm{CMB}}{T_\mathrm{408}\left(\Omega\right)-T_\mathrm{CMB}}\right)}}{\log{\left(\frac{230}{408}\right)}}.
\end{equation}

$T_\mathrm{230}\left(\Omega, t\right)$ is the same GSM base map at 230\,MHz, rotated as appropriate for the antenna location and the observing time, $t$. This was used as the base for the simulated sky so that the base map was close to, but not in, the band in which the 21-cm signal is located.

This spatially varying spectral index map was used in the place of a spatially uniform spectral index, as it is more physically realistic. Furthermore, non-uniformity in the spectral index introduces a change in the spatial distribution of brightness temperature on the sky with frequency that interacts with antenna chromaticity in non-trivial ways. This exacerbates the resulting chromatic distortion beyond what a simulation that assumes a uniform spectral index could capture, as is discussed in \cite{anstey21}. Therefore, we use a non-uniform spectral index in the modelling process to account for this.

Next, a 3-parameter, Gaussian, given by
\begin{equation}\label{eq:signal_model}
    T_\mathrm{signal}\left(\nu\right) = -A\exp{\left(-\frac{\left(\nu-f_\mathrm{0}\right)^{2}}{2\sigma^2}\right)},
\end{equation}
parameterised by the amplitude $A$, the centre frequency $f_\mathrm{0}$ and the standard deviation $\sigma$, is added to the data to simulate the 21-cm signal. Making the assumption that the 21-cm signal can be considered to be spatially uniform for the purpose of this experiment means that mock 21-cm signals can be linearly added to the simulated foreground data $T_\mathrm{simulated}\left(\nu\right)$. 

The resulting simulated data is then analysed using the REACH data analysis pipeline, to attempt to recover the input 21-cm signal. The REACH pipeline works by simulating an approximate sky model using parameterised spatially varying spectral indices, calculating what the resulting chromatic data would be using a provided antenna directivity, then fitting this to the input observational data using the nested sampling algorithm \texttt{PolyChord} \citep{handley15a,handley15b} to derive the spectral index parameter values. This process enables much of the chromatic distortion to be modelled as part of the foreground, allowing the 21-cm signal parameters to be derived for antennae that are not perfectly achromatic.

The approximate sky model is generated by subdividing the sky into a number of regions, $N$, and making the assumption that the spectral index is uniform within each. Dividing the sky into a greater number of regions results in a more detailed model that can theoretically describe the true sky more accurately, at the cost of requiring more parameters. As the fit is performed using nested sampling, the Bayesian evidence calculated by the nested sampling algorithm can be used to select the optimum number of regions based on which has the highest evidence.

This analysis can be repeated for a sweep over parameters of the input 21-cm signal. Analysing the resulting parameter fits for each will then give information about the range of potential 21-cm signals that the antenna would be able to detect, and how accurately it would do so. By performing this analysis for many antennae, it will indicate how well each will perform when used in a global 21-cm experiment. These results can then be used to guide the design of an antenna for such an experiment. A full analysis on an antenna can take of order several days to run, which does restrict how extensively this can be used. Therefore, it is most efficiently applied to aid in broad selection of a design, as is demonstrated here, or for comparing the effects of fine details in final designs, after other, faster optimisation processes have been used.

In order to cover a wide range of possible 21-cm signals, we ran this test with 21-cm signals of centre frequency $f_0$ = 80\,MHz, 110\,MHz, 140\,MHz and 170\,MHz and amplitudes, $A$ = 0.155\,K, 0.1\,K and 0.05\,K. In each case, the signal had a width of $\sigma$ = 15\,MHz. The test was also run on data to which no 21-cm signal was added.

For each simulated 21-cm signal with each antenna, the data was fit with two different models. One included a model of both the foreground and a 3-parameter Gaussian, as described in \Cref{eq:signal_model}, and the other was a foreground model on its own. The difference in evidence between these two fits quantifies the confidence in the presence of a signal in the data. For example, a difference of 3 between the log evidence of the model that includes a signal and the model that does not would mean there are betting odds of $\sim20:1$ in favour of a signal being present in the data.

This analysis was performed for foreground models using 7, 8, 9 and 10 sky regions. The optimum number can then be inferred from the Bayesian evidences. This range of $N$s was chosen as the evidence was found to peak in this range for most of the antennae used. Some of the more chromatic antennae used, however, do not peak in this range. If one of these were to be used in a real experiment, therefore, the analysis should be run at higher $N$. However, the range 7-10 is used in this analysis to allow like-for-like comparisons with the other antennae. Furthermore, the $N$ at which the evidence peaks or doesn't peak can give an indication as to the quality of the antenna for these experiments, by indicating how complex the foreground model must be to adequately account for the chromatic distortions. 

\section{Results}\label{sec:results}

\begin{figure*}
 \includegraphics[width=\textwidth]{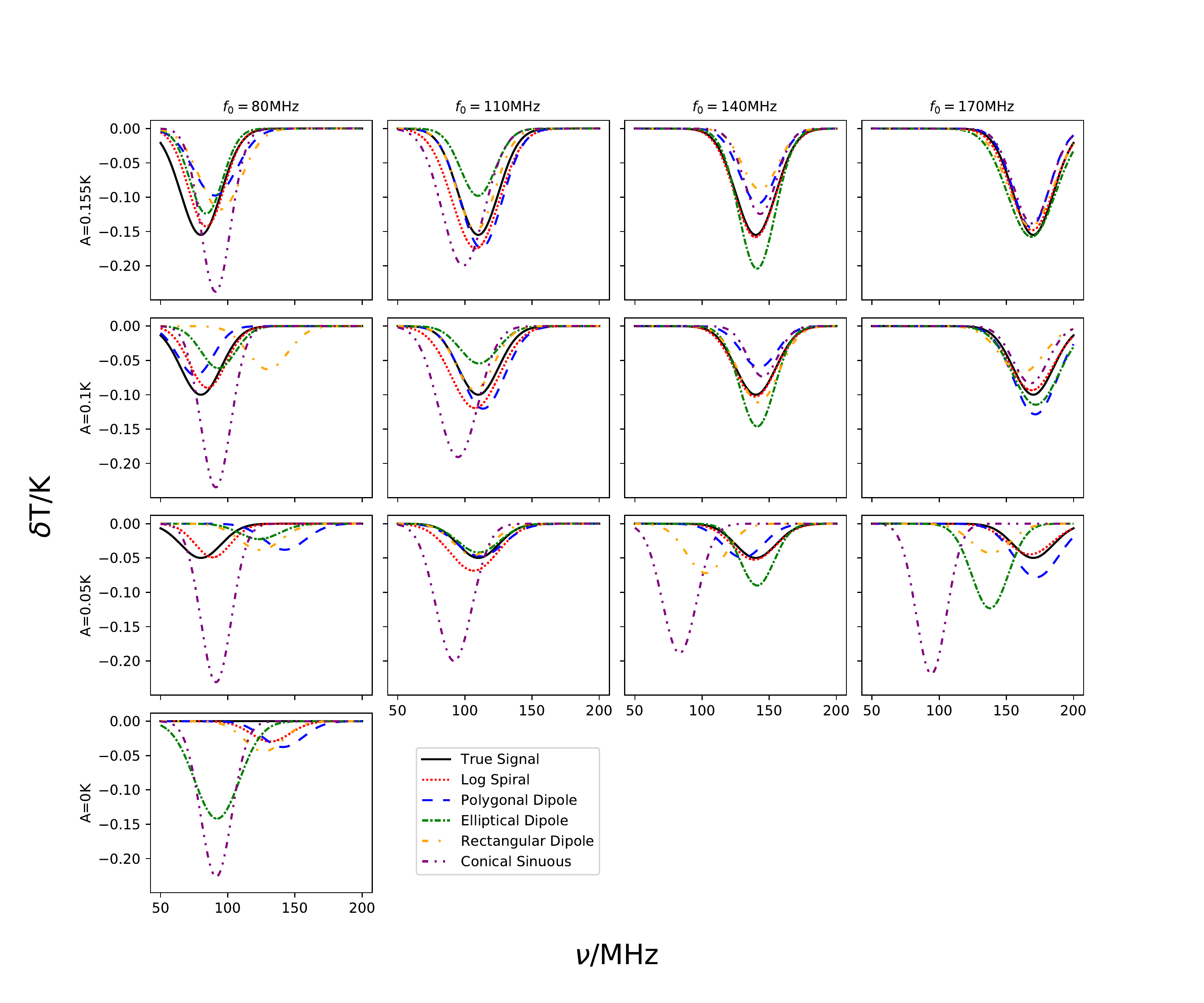}
 \caption{Comparison of the reconstructed signal models of the five antennae tested. In each case, the true signal inserted into the simulated data is shown in black. The coloured lines show the Gaussian signals given by the posterior averages of the fitted parameters shown in Figures \ref{fig:log_spiral_corner} - \ref{fig:con_sin_corner}. In each case the posterior of the model fit that gave the highest evidence is used. The numbers of regions $N$ that these correspond to are given in \Cref{tab:peak_N}.}
 \label{fig:signal_comparison}
\end{figure*}

\begin{figure*}
 \includegraphics[width=\textwidth]{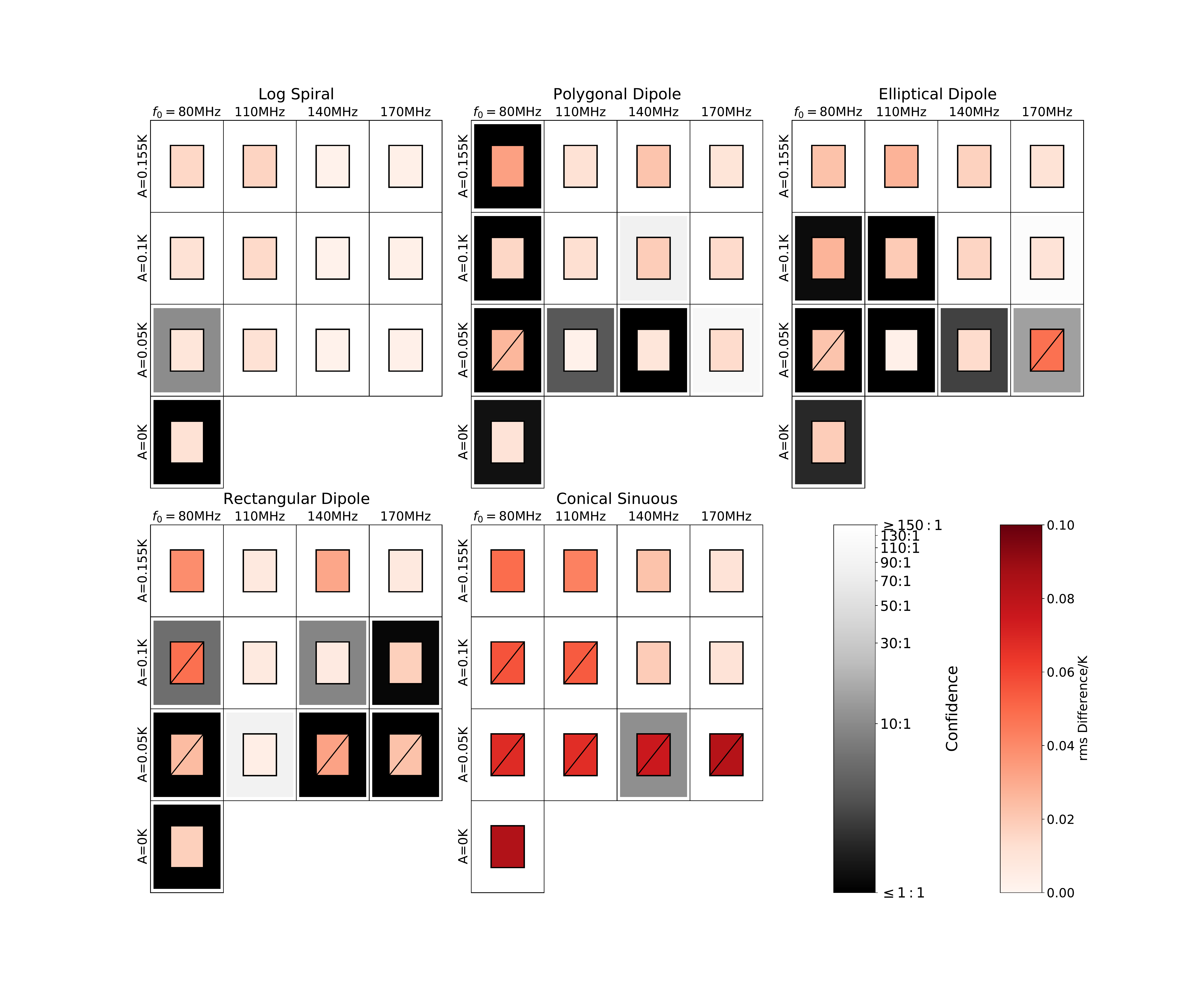}
 \caption{Plot of the signal detection confidence for each test, calculated from the difference between the evidences of a joint fit of a foreground and signal model together, $\mathcal{Z}_\mathrm{S+F}$, and a fit of a foreground model alone, $\mathcal{Z}_\mathrm{F}$, to the data in each case. `Confidence' describes the betting odds in favour of a signal being present, given by $e^{\log\left(\mathcal{Z}_\mathrm{S+F}\right) - \log\left(\mathcal{Z}_\mathrm{F}\right)}$. In each case, the models with the number of regions, $N$, that have the highest $\mathcal{Z}_\mathrm{S+F}$ and highest $\mathcal{Z}_\mathrm{F}$, as specified in \Cref{tab:peak_N}, are used. The confidence of a detection is given by the outer squares. The inner squares show a figure of merit of the recovered signal accuracy, given by the root-mean-square difference between the true signal inserted to the data and the associated posterior-average signal given in \Cref{fig:signal_comparison}. A line across the inner square indicates that the rms difference is greater than a third of the true signal amplitude.}
 \label{fig:evidence_compilation}
\end{figure*}

The abilities of each of the five antennae tested, a conical log spiral, inverted conical sinuous and rectangular-, polygonal- and elliptical-bladed dipoles, to recover each inserted signal are compared in \Cref{fig:signal_comparison} and \Cref{fig:evidence_compilation}. \Cref{fig:signal_comparison} shows, for each antenna and inserted signal, the Gaussian signal given by the mean of the fitted parameter posteriors, which are shown in Figures \ref{fig:log_spiral_corner} - \ref{fig:con_sin_corner}. The highest evidence $N$ of the four runs is used in each case, which are listed in \Cref{tab:peak_N}.

\Cref{fig:evidence_compilation} shows a summary of the confidence with which each input signal was identified by each antenna, focused on the marginal cases with betting odds of $\leq150:1$. This is calculated from the difference in evidence between a model that includes a signal and a model that doesn't. In each case, the model $N$ used is the one with the highest evidence of the four $N$s tested. These are listed in \Cref{tab:peak_N} and are not necessarily the same for models with and without a signal. It also shows a figure of merit of the accuracy of the recovered signals. This is given by the root-mean-square differences between the true signals inserted into the data, and the posterior average signals shown in \Cref{fig:signal_comparison}. More detailed plots of the signal posteriors for every model fit performed for each antenna are given in \Cref{appendix}.

In general, these results show that for all tested antennae, signals of larger amplitude and higher centre frequency are recovered more accurately and with greater confidence. The power law nature of foregrounds in this frequency range means sky temperatures at the lower end of the band can be more than an order of magnitude larger than those at the higher end of the band. The chromatic distortions that arise due to the coupling of these foregrounds to the chromaticity of the antenna are also correspondingly larger. It can therefore be expected that signals of larger amplitude and higher frequency are easier to detect as seen here, since the ratio of signal amplitude to chromaticity amplitude will be larger.

 The specific performance of each antenna will be discussed in the following sections.

\begin{table*}
    \caption{List of the number of regions $N$ that the sky was subdivided into that gave the highest evidence for each antenna and inserted signal tested for both the case of model of both foreground and signal and a model of the foreground alone.}
    \label{tab:peak_N}
    \centering
    \begin{tabular}{cc|cccc|cccc}
        \toprule
        Antenna & Signal Amplitude $A\mathrm{/K}$ & \multicolumn{8}{c}{\centering Signal Centre Frequency $f_0\mathrm{/MHz}$} \\
         & & 80 & 110 & 140 & 170 & 80 & 110 & 140 & 170\\  
         \midrule
         & & \multicolumn{4}{c}{\centering Foreground and Signal} & \multicolumn{4}{c}{\centering Foreground Only} \\
         \midrule
         \multirow{4}{1.2cm}{\centering Log Spiral} & 0.155 & 8 & 8 & 8 & 8 & 8 & 9 & 9 & 9\\ 
          & 0.1 & 8 & 8 & 8 & 8 & 8 & 9 & 9 & 9\\
          & 0.05 & 8 & 8 & 8 & 8 & 8 & 9 & 9 & 9\\ 
          & 0.0 & 8 &  &  &  & 8 &  &  &  \\ 
          \midrule
         \multirow{4}{1.2cm}{\centering Polygonal Dipole} & 0.155 & 9 & 10 & 9 & 8 & 9 & 9 & 10 & 10\\ 
          & 0.1 & 10 & 10 & 9 & 9 & 9 & 9 & 9 & 10\\
          & 0.05 & 9 & 8 & 9 & 9 & 9 & 8 & 9 & 10\\ 
          & 0.0 & 9 &  &  &  & 8 &  &  &  \\ 
          \midrule
         \multirow{4}{1.2cm}{\centering Elliptical Dipole} & 0.155 & 7 & 8 & 7 & 7 & 7 & 8 & 10 & 10\\ 
          & 0.1 & 7 & 8 & 7 & 8 & 7 & 8 & 9 & 10\\
          & 0.05 & 7 & 8 & 7 & 8 & 7 & 8 & 9 & 9 \\ 
          & 0.0 & 8 &  &  &  & 9 &  &  &  \\ 
          \midrule
         \multirow{4}{1.2cm}{\centering Rectangular Dipole} & 0.155 & 10 & 10 & 7 & 10 & 10 & 10 & 7 & 10\\ 
          & 0.1 & 10 & 10 & 10 & 10 & 10 & 10 & 7 & 10\\
          & 0.05 & 10 & 10 & 7 & 10 & 10 & 10 & 7 & 10\\ 
          & 0.0 & 10 &  &  &  & 10 &  &  &  \\ 
          \midrule
         \multirow{4}{1.2cm}{\centering Conical Sinuous} & 0.155 & 10 & 10 & 10 & 10 & 10 & 10 & 10 & 7 \\ 
          & 0.1 & 10 & 10 & 10 & 10 & 10 & 10 & 10 & 7\\
          & 0.05 & 10 & 10 & 10 & 10 & 10 & 10 & 10 & 10 \\ 
          & 0.0 & 10 &  &  & & 10 &  &  &  \\ 
   \bottomrule
    \end{tabular}
\end{table*}

\subsection{Log Spiral Antenna}\label{sec:log_spiral}
The results in \Cref{fig:signal_comparison} and \Cref{fig:evidence_compilation} show the log spiral antenna to be the most consistent at allowing 21-cm signals across the entire tested parameter space to be detected in the data. For this antenna, a signal very close to the true inserted signal is recovered in every case, with very high confidence. 

Even low amplitude signals with $A$ = 0.05\,K are also accurately and confidently identified, showing the log spiral antenna can detect even very small 21-cm signals. However, when no signal is present, this antenna also correctly measures no confidence in the presence of a signal.

\Cref{fig:log_spiral_corner} shows the signal parameter posteriors and model fit evidences for the log spiral antenna. It can be seen from these results that when a log spiral antenna is used for the simulated observations, the REACH data analysis pipeline is able to recover the signal parameters to within around two standard deviations at most for all the signals tested. Signals at higher frequencies tend to be identified more accurately, but the signal is correctly identified at low frequencies as well. This is reasonable, as a signal of larger amplitude in a frequency range where the foregrounds and chromatic distortions are smaller should be easier to detect, independent of the specific antenna design.

The least accurate posteriors for every inserted signal are those with $N=7$. However, these are also the results with the lowest evidences by a significant margin in all cases. Therefore, the models with $N$ in which the signal is more successfully reconstructed are also strongly preferred, meaning the signal can be considered to have been accurately identified in each case, despite the less accurate $N=7$ fits.

The evidence difference between models with and without a signal, shown by blue lines in \Cref{fig:log_spiral_corner}, can be seen to decrease with decreasing signal amplitude. This means the nested sampling pipeline's confidence in the presence of a signal decreases as the signal decreases in magnitude, as could be reasonably expected. However, when a signal is actually present in the data, the minimum value for this difference is $\sim 2.5$, for $f_0 = 80\mathrm{\,MHz}$ and $A=0.05\mathrm{\,K}$. It is reasonable for this to be the least confident fit, as it is the smallest signal at the centre frequency where the foregrounds are largest. However, a difference in log evidence of $2.5$ corresponds to betting odds of $\sim 12:1$ in favour of a signal being present. Therefore, the analysis still detects this signal with a high degree of confidence, and all other signals are identified to even higher certainty.

When no signal is present in the data, the $f_0$ and $\sigma$ parameters can be seen to be much less constrained than when a signal was present and the $A$ parameter posterior is around zero. Furthermore, the evidence difference is negative for all $N$, showing a statistical preference for no signal being present. This all demonstrates that the lack of a signal can be correctly identified when a log spiral antenna is used. 

\subsection{Polygonal Dipole Antenna}\label{sec:hex_dipole}
The polygonal dipole antenna is the next best performing antennae, although it is closely followed by the elliptical dipole. They perform almost as well as the log spiral, apart from when the signal has a very low amplitude or is at low frequencies.

For the case of the polygonal dipole, \Cref{fig:signal_comparison} and \Cref{fig:evidence_compilation} show that it is able to accurately and confidently detect most of the signals. However, the reconstructed signals are least accurate when the true signal has $f_0 = 80\mathrm{\,MHz}$, and there is no statistical confidence in the presence of a signal in the data in these low frequency cases as well. There is also an additional low amplitude signal, specifically, when $f_0 = 140\mathrm{\,MHz}$ and $A=0.05\mathrm{\,K}$, where the signal is also only very tenuously detected.

Furthermore, as for the log spiral antenna, using a polygonal dipole antenna allows it to be correctly identified when no signal is present in the data.

These results are reinforced by the posteriors in \Cref{fig:hex_dipole_corner}. These model fits are less consistent with changing $N$ than for the log spiral. However, as before, the highest evidence model should be considered as the best result in each case. If only the highest evident $N$ is considered, the results are similar to those of the log spiral antenna, with the signal parameters recovered to within 2 standard deviations in most cases and the posteriors more accurate to the true parameters for signals at higher frequencies. 

As discussed previously, the main distinctions from the results of the log spiral are those where the signal has $f_0 = 80\mathrm{\,MHz}$, and that with $f_0 = 140\mathrm{\,MHz}$ and $A=0.05\mathrm{\,K}$, which have much broader, less constrained posteriors, and correspondingly lower confidence. In addition, for all injected signals, the statistical confidence in the recovered signals for this polygonal dipole antenna are significantly less than those of the log spiral antenna, although still high in most cases. It can also be noted that the peak $N$s for this antenna are almost all higher than those for the log spiral, meaning a more complex model is needed to account for it's chromatic structure. This is in agreement with the higher chromaticity of the polygonal dipole relative to the log spiral seen in \Cref{fig:waterfall_plots}.

Overall, a polygonal dipole antenna is less effective at producing accurate 21-cm signal detections than a log spiral antenna, especially for the signals with centre frequencies of $80$\,MHz and amplitudes of $0.05$\,K. The statistical confidences in these detections are also lower than when a log spiral antenna is used. However, it still all other 21-cm signals tested to be accurately and confidently identified. These results, and those for all antennae tested, could theoretically be improved further by fitting multiple data sets at different observing times simultaneously, as opposed to a single time-integrated data set. This possibility will be discussed further in \Cref{sec:conclusions} as potential future work.

\subsection{Elliptical Dipole Antenna}\label{sec:elliptical_dipole}
The performance of the elliptical antenna is very similar to that of the polygonal dipole, with a few key differences. From \Cref{fig:signal_comparison} and \Cref{fig:evidence_compilation}, it can be seen that the elliptical dipole is able to accurately and confidently detect the highest amplitude signal at $f_0 = 80\mathrm{\,MHz}$, which the polygonal dipole could not. However, all other $f_0 = 80\mathrm{\,MHz}$ signals are not detected with confidence by this antenna, as with the polygonal dipole. 

In addition, the elliptical dipole is not able to detect with confidence the signal at $f_0 = 110\mathrm{\,MHz}$ and $A=0.1\mathrm{\,K}$, which was detected by the polygonal dipole. Furthermore, \Cref{fig:signal_comparison} shows that the reconstructed low amplitude $A=0.05\mathrm{\,K}$ signals are all noticeably less accurate than those of the polygonal dipole. 

It can also be seen that in the case where no signal was present in the data, the model fit produces a false positive signal detection of $A\approx0.14\mathrm{\,K}$ and $f_0\approx90\mathrm{\,MHz}$, albeit with a very low confidence of $\sim 2:1$ that such a signal is actually present.

The parameter posteriors in \Cref{fig:elliptic_dipole_corner} provide some additional detail. As with the polygonal dipole, the less confident detections have correspondingly less constrained posteriors. It can be seen that this elliptical dipole also has a greater range of signals for which it shows these poorly constrained posteriors than the polygonal dipole did. 

Furthermore, as previously mentioned, for the case of $A=0.05\mathrm{\,K}$ and $f_0=170\mathrm{\,MHz}$, the highest evidence model is a relatively confident detection,with betting odds of $\sim14:1$ in favour of the signal being present, at a centre frequency of $\sim 140\mathrm{\,MHz}$, which is lower than the true signal. The implications of such confident false signals will be discussed in \Cref{sec:con_sin}.

In general, this dipole performs similarly to the polygonal dipole, although with a slightly lower range of signals it is capable of detecting, especially at low frequencies. However, this is in spite of the fact that this antenna appears superficially less chromatic that the polygonal dipole. It has a slightly lower range of variation of $A_\mathrm{chromaticity}\left(t,\nu\right)$, and generally achieves an evidence peak at lower $N$, less complex models. This slightly lower performance may instead be due to the fact that, although it has a lower extent of change in $A_\mathrm{chromaticity}\left(t,\nu\right)$, the changes are much steeper and more abrupt than for the polygonal dipole. 
This antenna also carries the risk of making a false-positive signal detection owing to additional residual chromaticity beyond what the model can capture, although the low confidence can allow them to be identified as such.

\subsection{Rectangular Dipole Antenna}\label{sec:square_dipole}
The results of a rectangular bladed dipole antenna are similar to those of the elliptical dipole. The primary differences, however, as can be seen in \Cref{fig:square_dipole_corner}, are that all fits for signals of $A\leq0.1\mathrm{\,K}$ show relatively wide, poorly constrained posteriors for the maximum evidence $N$ case. For the highest amplitude and higher centre frequency cases, the signal is still correctly and confidently detected. However, it is much less precise than the previously discussed antennae. The rectangular dipole is also unable to confidently detect many signals of very low amplitude or centre frequency, as for the other dipoles.

Furthermore, it can be seen in \Cref{fig:square_dipole_corner} that, while the previous three antenna all showed a peak in evidence within the $N=7$-$10$ range, which is why this range was chosen, the rectangular dipole has not yet reached a peak in evidence by $N=10$. This is indicative that this antenna has a more chromatic distortion pattern than the previous ones.

\subsection{Conical Sinuous Antenna}\label{sec:con_sin}
The conical sinuous antenna can also be seen as the least effective at enabling the detection of the 21-cm signal, especially when the signal amplitude is low. From \Cref{fig:signal_comparison}, it can be seen that for all inserted signals of $A \leq 0.05\mathrm{\,K}$ and $f_0 \leq 110\mathrm{\,MHz}$, the signal is not correctly or accurately reconstructed. However, \Cref{fig:evidence_compilation} shows that these incorrect signals are still detected with a high degree of confidence. This is also the case when no signal is present in the data. A few false detections similar to these were also seen in the results of the elliptical and rectangular bladed dipoles, but they are most prominent for the conical sinuous antenna. 

This is indicative that this antenna is too chromatic for the REACH data analysis foreground modelling technique used here to fully model the chromatic distortion it introduces, resulting in residual chromatic structure that the signal model is fit to. False detections such as these are discussed in \citep{anstey21} and primarily occur in more strongly chromatic antennae for foreground models with $N$ below the optimum value, as is the case for the conical sinuous antenna here. Increasing $N$ towards the optimum value will eliminate these false positive detections.

The fact that the conical sinuous antenna is able to detect the fewest signals of the five antennae tested, and has one of the more difficult to model chromatic structures, given that it has not yet reached a peak by $N=10$, is not something that could be directly determined from the $A_\mathrm{chromaticity}\left(t,\nu\right)$ in \Cref{fig:waterfall_plots}. The conical sinuous antenna actually has the second smallest range of change after the log spiral antenna. However, upon deeper analysis, we find that it is actually a poor design for a global 21-cm experiment.

\section{Discussion and Conclusions}\label{sec:conclusions}
Chromatic distortion due to the antenna can have a significant impact on global 21-cm experiments. Whilst this can be mitigated somewhat by performing modelling and corrections, as the REACH data analysis pipeline does, the effect is very difficult to remove entirely without near-perfect prior knowledge of both the foregrounds and the antenna beam. As information of the required accuracy and precision is not currently available in practice, it is therefore very important, when designing such an experiment, to consider exactly what effect the chromaticity will have, and to use an antenna whose chromaticity can be corrected for or modelled without distorting the signal.

Some information can be obtained by measuring the change in the antenna pattern directly. However, it is very difficult to tell from this directly whether an antenna is suitable for a global 21-cm experiment, as the interaction of distortions with the 21-cm signal is very non-trivial. Certain patterns can be easier to correct for, or more spectrally distinct form the signal than others.

We therefore propose here a method of simulating a global 21-cm experiment with a given antenna for many different input signals, in order to produce a more comprehensive analysis of which signals a given antenna can and cannot detect, and how accurately it will do so. We demonstrate this analysis on five different antennae that were considered for use in the REACH experiment. 

A few general trends were observed in all five antennae tested. Firstly, as could be trivially expected, all antennae were less accurate at detecting signals of lower amplitude, and yielded less confidence in the presence of such signals in the data. All antennae were also less accurate and provided less confidence in identifying signals at lower frequencies. This is due to the fact that the foregrounds can be up to an order of magnitude brighter around $\sim 50\mathrm{\,MHz}$ than they are at $\sim 200\mathrm{\,MHz}$ due to their power law nature. This means the ratio of signal amplitude to foreground amplitude and chromatic distortion amplitude is larger at these frequencies, making the signal harder to detect.

The log spiral antenna was found to perform the best of the five tested. Using this antenna in the simulations meant signals across the entire $50-200\mathrm{\,MHz}$ band could be accurately recovered from the data even when the signal amplitude was very low, at $A=0.05\mathrm{\,K}$. Furthermore, when the log spiral antenna was used, it could also be correctly determined when no signal was present in the data. This is in line with the fact that the log spiral antenna had both the lowest magnitude of chromatic distortion of the five antennae, and a very regular oscillatory pattern of distortion, which is much easier to model than other patterns.

The conical sinuous antenna, however, performed significantly worse than all the other antennae tested. When this antenna was used, only the largest amplitude signals ($A=0.155\mathrm{\,K}$) could be identified in the data, and often with biased amplitudes. For lower amplitude signals, a highly inaccurate signal was identified instead. This behaviour was not necessarily apparent from the chromaticity of the beam, as the conical sinuous antenna had the lowest degree of chromaticity after the log spiral antenna. This is likely because, although the degree of chromaticity is relatively low, the pattern is highly irregular and produces distortions that are spectrally very similar to the signal, which makes the signal difficult to accurately distinguish.

The three dipoles tested had mostly similar chromatic patterns, and were observed to perform roughly equivalently as well. They were able to detect most large amplitude 21-cm signals, but began to have difficulty detecting low amplitude and low frequency signals. Of these three dipoles, however, the polygonal-bladed dipole was found to perform the best. It was able to confidently detect a wider range of lower amplitude signals than the other two dipoles and also reconstructed a more accurate signal in a greater number of cases.

There are however, a number of limitations to this method that should be taken into account. In particular, the results presented here quantify signals which each antenna can or cannot detect under ideal conditions, in which the antenna beam is known perfectly, on entirely flat terrain with no RFI, soil emissions, or other complications that occur in practice. Therefore, an antenna producing a confident and correct detection of a given signal in the simulations presented here does not guarantee that the antenna would be able to detect that signal in a real experiment. However, a successful detection under simulated ideal conditions is a necessary prerequisite for a detection in practice, so these simulations are still valid in comparing the effectiveness of antennae and informing the design process. Furthermore, this analysis was only performed on data of 1 hour of integration over a specific time frame of 8.12 to 9.12 hrs LST. It is shown in \citet{anstey21} that the REACH pipeline can perform slightly differently for different observation and integration times. Therefore, this analysis could be expanded to compare the performance of antennae for different times. Another way to extend this analysis would be to analyse a more finely gridded range of simulated signals.

In addition, the pipeline used in this analysis works by fitting a single parameterised model to a single, time-integrated data set. If instead, multiple data sets at different observing times were fit simultaneously to multiple corresponding models that all used the same parameters, the foreground and chromaticity structure of the antenna used could potentially be determined even more accurately, which would result in further improvements in the performance of the antennae by this process. This extension will be considered in future work.

Overall, this analysis demonstrates that, of the five prototypes tested here, the log spiral and polygonal-bladed dipole antennae are the best suited for use in a global 21-cm experiment. When observational data sets from these antennae are analysed using the REACH pipeline, the chromaticity distortions can be modelled accurately enough for a wide range of 21-cm signals to be reliably detected. 

This is one of the factors that lead to the choice by the REACH collaboration to use these two antennae for the experiment. However, there are also many other metrics that need to be considered when selecting an antenna for a global 21-cm experiment. For example, although the log spiral antenna was seen to detect a wider range of signals more reliably and confidently than the dipole antennae tested, it is also much larger and harder to construct in practice than dipole antennae. A full consideration of the process of selecting and optimising the REACH antenna design and the various figures of merit used in the selection, such as S11 and ease of construction, are discussed in detail in Cumner et al. (in prep.) The technique discussed here was also applied again in the optimisation process of these two antennae, which will also be discussed in greater detail in this paper.

\section*{Acknowledgements}
We would like to thank Quentin Gueuning for providing the electromagnetic simulations and image of the log spiral antenna.
Dominic Anstey and Eloy de Lera Acedo were supported by the Science and Technologies Facilities Council. John Cumner was supported by EPSRC iCASE in partnership with BT. Will Handley was supported by a Royal Society University Research Fellowship. We would also like to thank the Kavli Foundation for their support of REACH. The posterior plots in this paper were produced using \texttt{anesthetic} \citep{anesthetic}.

\section*{Data Availability}
The data underlying this article is available on Zenodo, at https://dx.doi.org/10.5281/zenodo.4672634.




\bibliographystyle{mnras}
\bibliography{bibliography} 



\appendix

\section{Model posterior details}\label{appendix}

The posteriors of every model fit are given in the following plots. For each antenna, we give the posteriors of the three parameters of the Gaussian signal model ($f_0$, $A$ and $\sigma$) for each of the four $N$ values tested (7, 8, 9 and 10). These posteriors are given for every inserted signal.

\begin{figure*}
    \begin{turn}{-90}
    \begin{minipage}{7.5in}
 \includegraphics[scale=0.34]{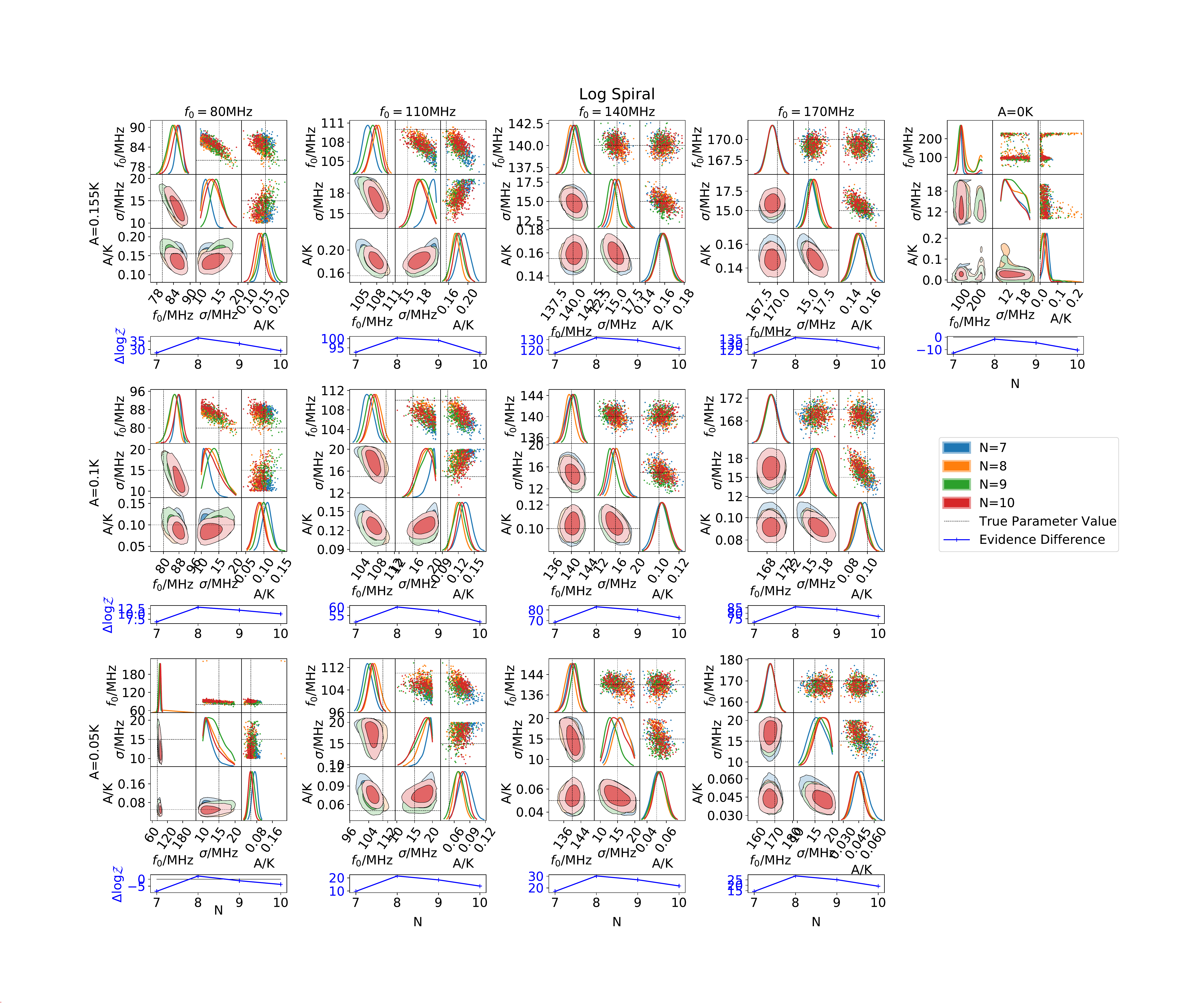}
 \end{minipage}
 \end{turn}
  \caption{Results of attempting to detect a range of 21-cm signals using the REACH data analysis pipeline in simulated observational data of a conical log spiral antenna. Each subplot shows the results for a different simulated 21-cm signal in the data, changing the centre frequency $f_0$ along the horizontal axis and the amplitude $A$ along the vertical axis. In each case the signal has a width of $\sigma$ = 15\,MHz The rightmost plot shows the results when no signal is added to the simulated data. In each subplot, the upper plot shows the 2D and 1D marginalised posteriors, and the marginalised posterior samples, of the three Gaussian signal parameters recovered in the model fitting. The posteriors of the foreground and noise parameters are not shown, as they are not equivalent across the four numbers of regions and so cannot be compared. The posteriors in blue, orange, green and red show the posteriors when a 7 region, 8 region, 9 region and 10 region foreground is used, respectively. The contours are at 1 and 2 standard deviations. The dashed black lines mark the true parameter values of the simulated signal inserted in each case. In the lower plot of each subplot, the solid blue line shows the difference in log evidence between the fits whose posteriors are shown, for each number of regions, $N$, and an equivalent fit of a model of the foreground alone, with no signal model, using a number of regions as specified in \Cref{tab:peak_N} that gave the highest evidence. This quantifies the confidence of the pipeline in the presence of the signal in the data. The solid black line marks where the evidence difference is zero. If $\Delta\log\mathcal{Z}>0$, the presence of a signal is statistically favoured, and if $\Delta\log\mathcal{Z}<0$, the absence of a signal is favoured.}
   \label{fig:log_spiral_corner}
\end{figure*}

\begin{figure*}
    \begin{turn}{-90}
    \begin{minipage}{9in}
 \includegraphics[width=\textwidth]{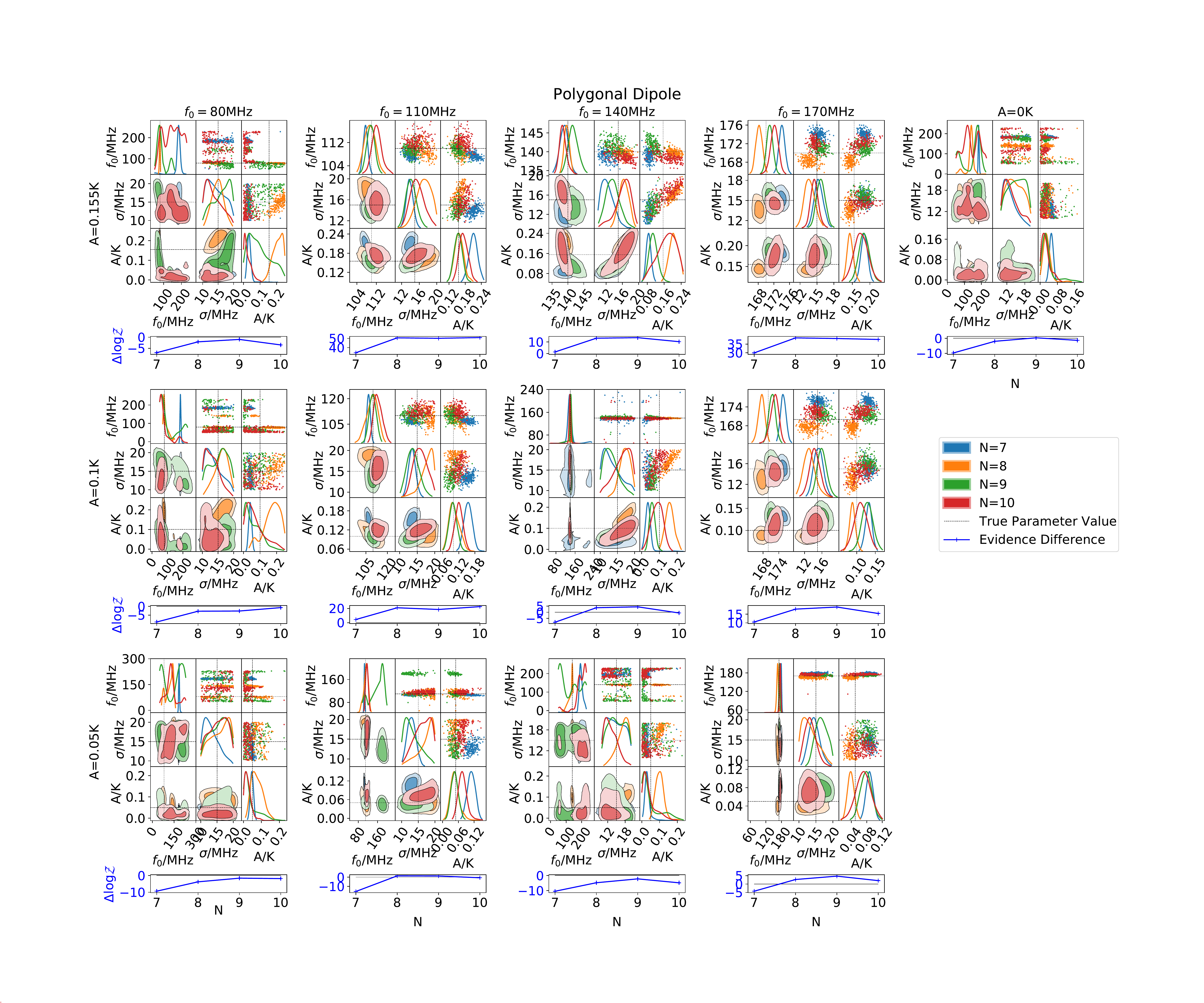}
 \end{minipage}
 \end{turn}
  \caption{Equivalent to \Cref{fig:log_spiral_corner}, but for observational data simulated using a polygonal dipole antenna.}
   \label{fig:hex_dipole_corner}
\end{figure*}

\begin{figure*}
    \begin{turn}{-90}
    \begin{minipage}{9in}
 \includegraphics[width=\textwidth]{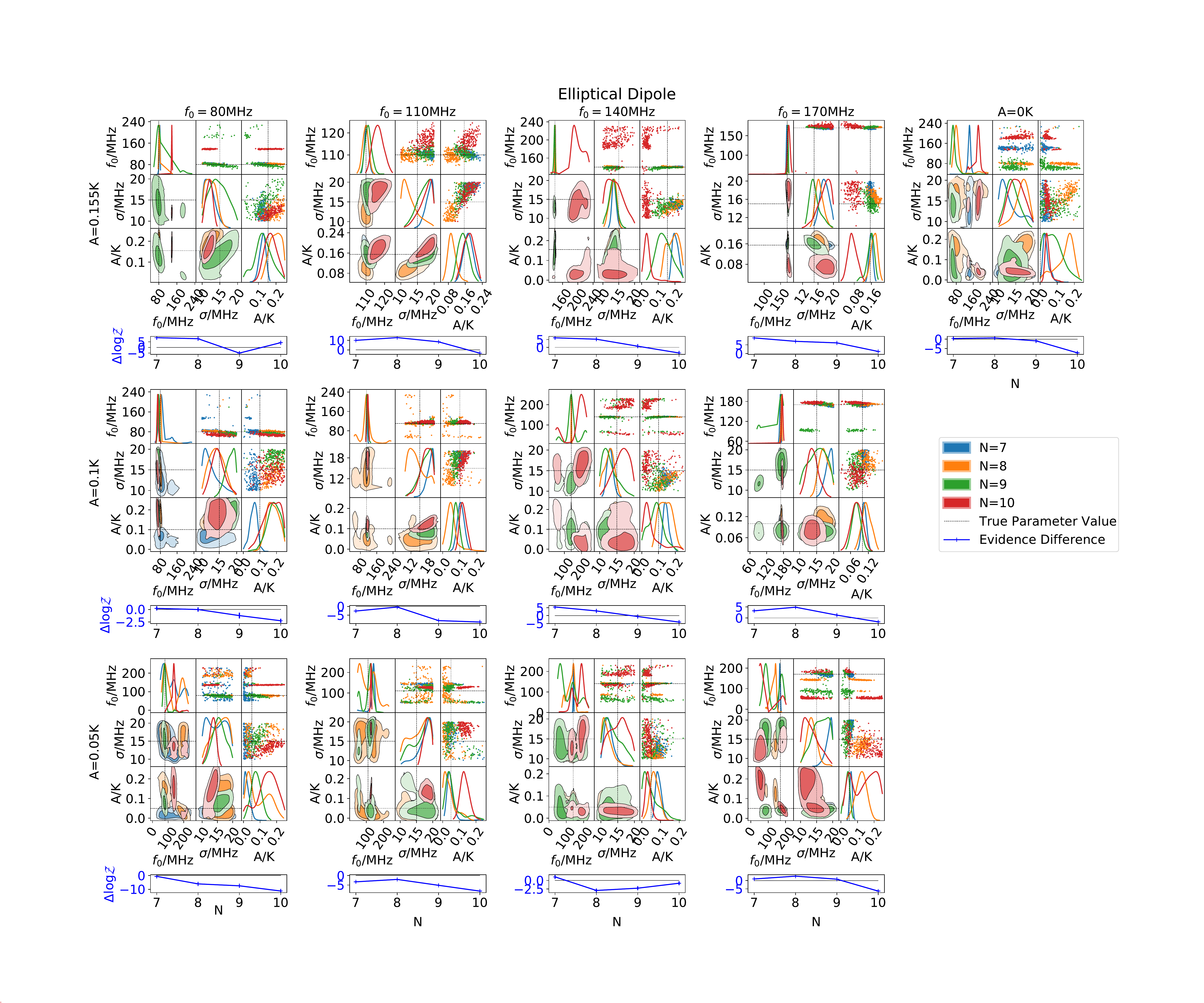}
 \end{minipage}
 \end{turn}
  \caption{Equivalent to \Cref{fig:log_spiral_corner}, but for observational data simulated using an elliptical dipole antenna.}
   \label{fig:elliptic_dipole_corner}
\end{figure*}

\begin{figure*}
    \begin{turn}{-90}
    \begin{minipage}{9in}
      \includegraphics[width=\textwidth]{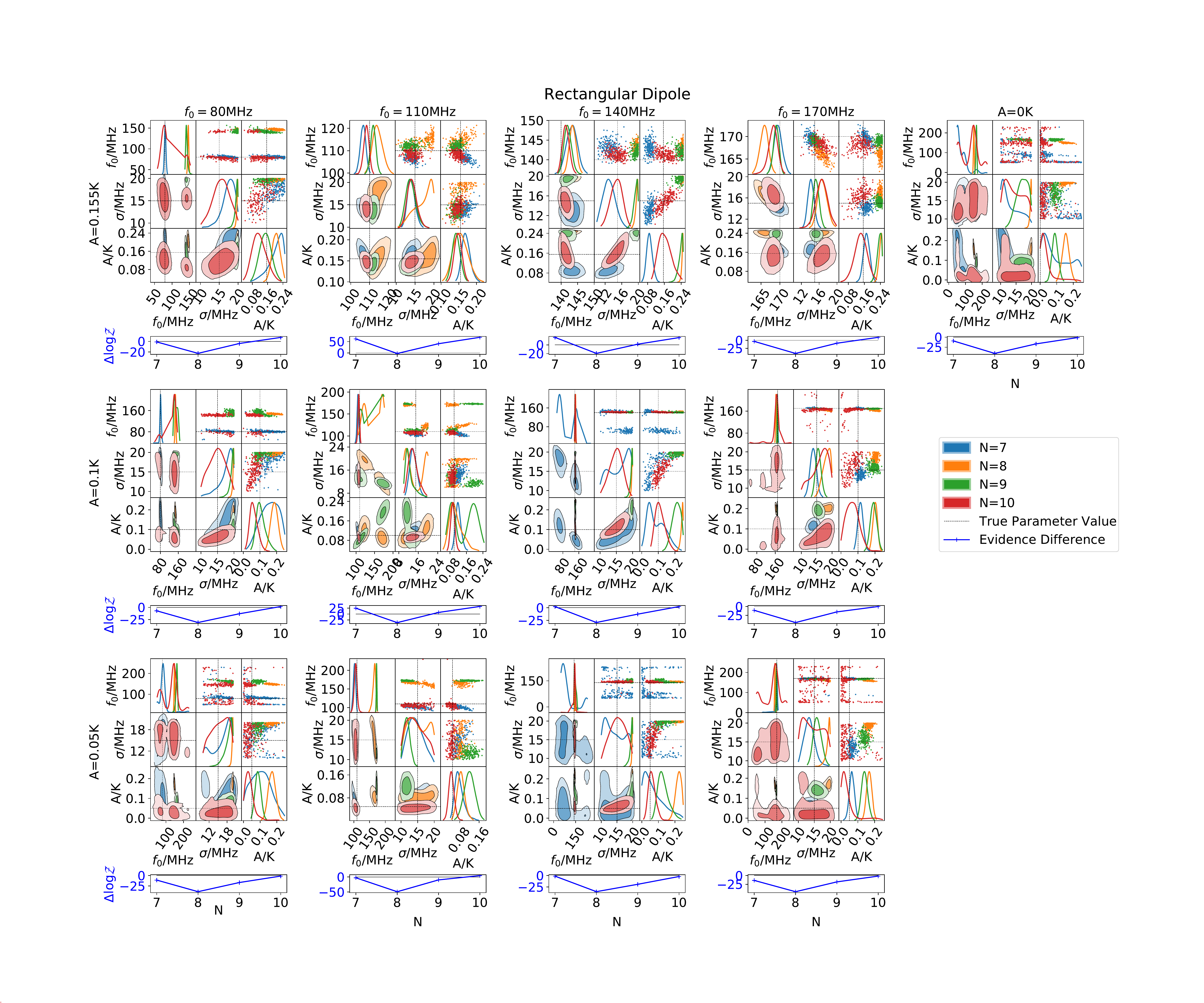}
 \end{minipage}
 \end{turn}
   \caption{Equivalent to \Cref{fig:log_spiral_corner}, but for observational data simulated using a rectangular dipole antenna.}
 \label{fig:square_dipole_corner}
\end{figure*}

\begin{figure*}
    \begin{turn}{-90}
    \begin{minipage}{9in}
 \includegraphics[width=\textwidth]{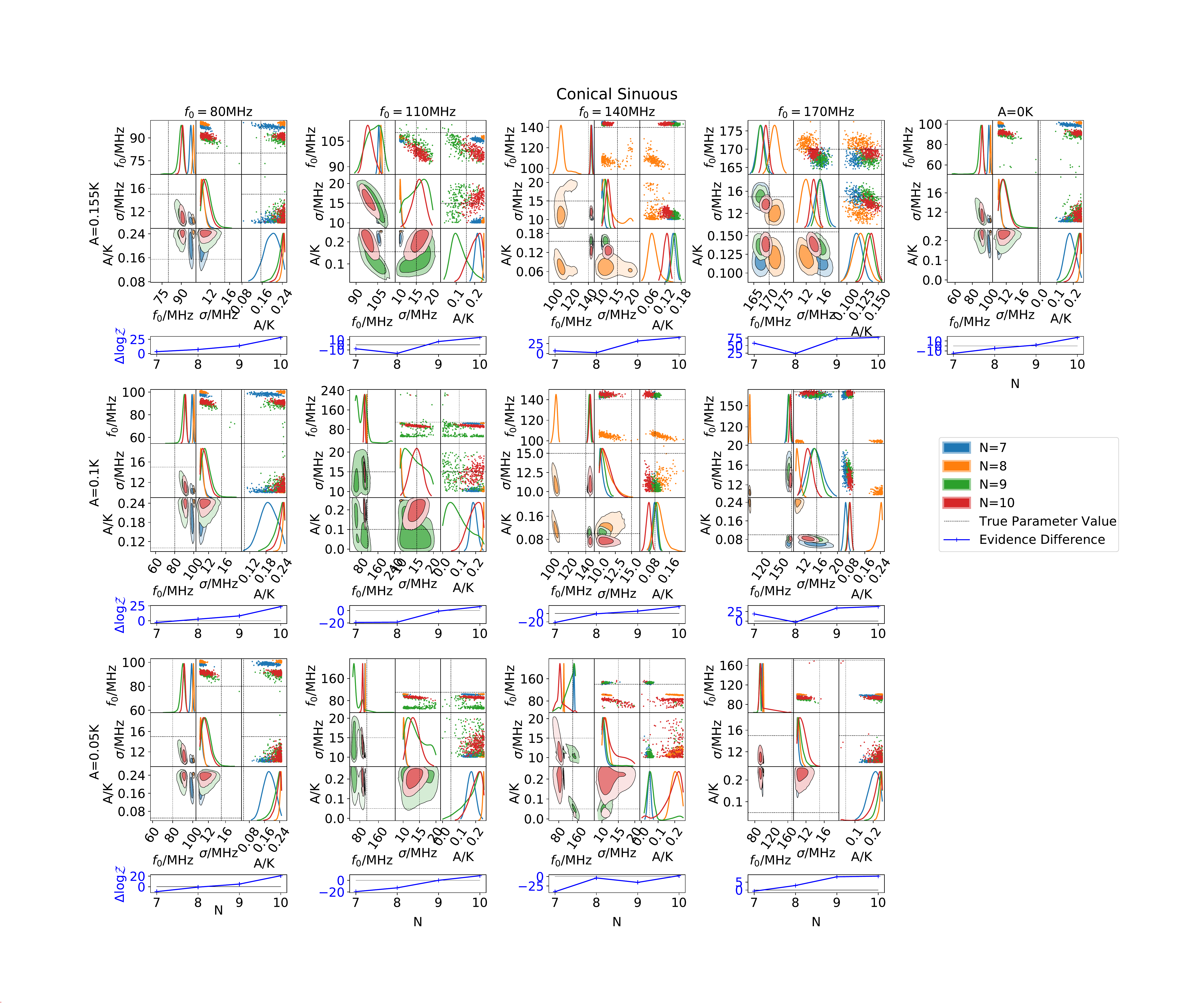}
 \end{minipage}
 \end{turn}
  \caption{Equivalent to \Cref{fig:log_spiral_corner}, but for observational data simulated using a conical sinuous antenna.}
 \label{fig:con_sin_corner}
\end{figure*}


\bsp	
\label{lastpage}
\end{document}